\documentclass[prd,onecolumn,nofootinbib]{revtex4}
\usepackage{bm,amsmath,amssymb}

\begin{document}

\noindent
Physical Review D {\bf 89} (2), 027501 (2014)\\

\title{Schwinger's variational principle in Einstein--Cartan gravity}
\author{Nikodem Pop{\l}awski}

\altaffiliation{NPoplawski@newhaven.edu}
\affiliation{Department of Physics, University of New Haven, West Haven, CT, USA}

\begin{abstract}
By applying Schwinger's variational principle to the Einstein--Cartan action for the gravitational field, we derive quantum commutation relations between the metric and torsion tensors.
\end{abstract}

\maketitle

The field equations of a classical theory are obtained in Hamilton's principle of stationary action by varying the action with respect to every variable $\phi$ and equating such a variation to zero for an arbitrary variation $\delta\phi$.
Equivalently, we calculate the variation of the Lagrangian density in the action integral and perform integration by parts.
Then we apply the Gauss--Stokes theorem and assume $\delta\phi=0$ on the boundary of the integration domain.
Finally, we equate to zero the coefficients of $\delta\phi$ inside the integration domain \cite{Schr}.

In a quantum theory, Hamilton's principle is generalized to Schwinger's variational principle, according to which the variation of the transition amplitude between an initial state $|\alpha_i\rangle$ and a final state $|\alpha_f\rangle$ is equal to $i/\hbar$ times the matrix element connecting the two states of the variation $\delta I$ of the action integral $I$ \cite{Schwi,var}:
\begin{equation}
\delta\langle\alpha_f|\alpha_i\rangle=\frac{i}{\hbar}\langle\alpha_f|\delta I|\alpha_i\rangle.
\label{Schw1}
\end{equation}
If Schwinger's principle is applied to variations that vanish on the boundary of the integration domain, the left-hand side of (\ref{Schw1}) is equal to zero.
Since the states $|\alpha_i\rangle$ and $|\alpha_f\rangle$ are arbitrary, the operator $\delta I$ vanishes, giving the principle of stationary action in an operator form.
If Schwinger's principle is applied to variations that vanish only for the initial state, the effect of the variation in the Schr\"{o}dinger picture is to apply a unitary operator $U=1+(i/\hbar)\delta I$ to the adjoint final state, which is equivalent to applying the Hermitian conjugate $U^\dagger$ to the final state, $|\alpha_f\rangle\rightarrow U^\dagger|\alpha_f\rangle$.
In the Heisenberg picture, the effect of this variation is to apply to any operator $O$ acting on the final state a unitary transformation $O\rightarrow UOU^{-1}=O-(i/\hbar)[O,\delta I]$, where square brackets denote a commutator.
Schwinger's variational principle is thus equivalent to \cite{Schwi,var,qft}
\begin{equation}
\delta O=-\frac{i}{\hbar}[O,\delta I].
\label{Schw2}
\end{equation}

For a particle in quantum mechanics, we substitute a coordinate operator $q_i$ for $O$ and use the formula for the action as a function of the coordinates and time at the final state, $\delta I=\delta\int Ldt=\sum p_j\delta q_j-H\delta t$, where $L$ is the Lagrangian, $p_j=\partial L/\partial\dot{q}_j$ denotes the generalized momentum corresponding to a generalized coordinate $q_j$, and $H=\sum p_j\dot{q}_j-L$ is the Hamiltonian \cite{LL1}.
The term $\delta I$ is the variation of the action at the boundary of the integration domain (over time).
If we consider a variation in which only the coordinates $q_i$ are changed, then $\delta I=\sum p_j\delta q_j$ and Schwinger's principle (\ref{Schw2}) gives $\delta q_i=-(i/\hbar)[q_i,\sum p_j\delta q_j]$, which is satisfied if $[q_i,q_j]=0$ and $[q_i,p_j]=i\hbar\delta_{ij}$.
By adding to the action $I$ a quantity $-\sum p_j q_j$, we bring its variation to $\delta I=-\sum q_j\delta p_j$.
Schwinger's principle (\ref{Schw2}) gives then $\delta p_i=(i/\hbar)[p_i,\sum q_j\delta p_j]$, which is satisfied if $[p_i,p_j]=0$ and $[p_i,q_j]=-i\hbar\delta_{ij}$.
Schwinger's principle therefore gives the canonical commutation relations of quantum mechanics \cite{Schwi,var,qft}.

For the electromagnetic field, the Lagrangian density is equal to $\mathfrak{L}=-(1/4)F_{\mu\nu}F^{\mu\nu}\sqrt{-g}$, where $F_{\mu\nu}=\partial_\mu A_\nu-\partial_\nu A_\mu$ is the electromagnetic field tensor, $A_\mu$ is the electromagnetic potential, and $g=\mbox{det}(g_{\mu\nu})$ is the determinant of the metric tensor $g_{\mu\nu}$ \cite{LL2,Niko}.
The action is thus equal to $I=-(1/4c)\int F_{\mu\nu}F^{\mu\nu}\sqrt{-g}d\Omega$, where $d\Omega$ is an element of the four-volume.
The variation of the actions is given by $\delta I=-(1/c)\int F^{\mu\nu}\partial_\mu\delta A_\nu\sqrt{-g}d\Omega=-(1/c)\int\partial_\mu(F^{\mu\nu}\sqrt{-g}\delta A_\nu)d\Omega+(1/c)\int\partial_\mu(F^{\mu\nu}\sqrt{-g})\partial A_\nu d\Omega=-(1/c)\oint F^{\mu\nu}\sqrt{-g}\delta A_\nu df_\mu+(1/c)\int\partial_\mu(F^{\mu\nu}\sqrt{-g})\partial A_\nu d\Omega$, where $df_\mu$ is an element of the closed hypersurface surrounding the integration four-volume.
We can consider a volume hypersurface $df_\mu=\delta^0_\mu dV$, where $dV$ is an element of the volume.
We also use $g=-g_{00}\gamma$, where $\gamma=\mbox{det}(\gamma_{\alpha\beta})$ is the determinant of the spatial metric tensor $\gamma_{\alpha\beta}=-g_{\alpha\beta}+g_{0\alpha}g_{0\beta}/g_{00}$ \cite{LL2,Niko}.
The hypersurface term, which should be used in Schwinger's principle (\ref{Schw2}), is thus $\delta I=-(1/c)\int\sqrt{g_{00}}F^{0\beta}\delta A_\beta\sqrt{\gamma}dV$.
This principle gives $\delta A_\alpha({\bf x},t)=(i/\hbar c)[A_\alpha({\bf x},t),\int(\sqrt{g_{00}}F^{0\beta}\delta A_\beta\sqrt{\gamma})({\bf x}',t)dV']$, which is satisfied if $[A_\alpha({\bf x},t),g_{\mu\nu}({\bf x}',t)]=[A_\alpha({\bf x},t),A_\beta({\bf x}',t)]=0$ and $[A_\alpha({\bf x},t),(\sqrt{g_{00}}F^{0\beta}\sqrt{\gamma})({\bf x}',t)]=-i\hbar c\delta^\beta_\alpha{\bm\delta}({\bf x}-{\bf x}')$.
Defining the spatial vector of the electric field, $E^\alpha=-\sqrt{g_{00}}F^{0\alpha}$ (such a vector appears in the Maxwell equations in curved space) \cite{LL2}, gives $[A_\alpha({\bf x},t),E^\beta({\bf x}',t)]=i\hbar c\delta^\beta_\alpha{\bm\delta}({\bf x}-{\bf x}')/\sqrt{\gamma}({\bf x},t)$, which generalizes to curved space the canonical commutation relations of electrodynamics \cite{Schwi,qft}.

For the gravitational field, we use the Einstein--Cartan theory of gravity, which naturally extends the general theory of relativity to matter with intrinsic angular momentum (spin) \cite{Niko,SK}.
In this theory, the affine connection $\Gamma^{\rho}_{\mu\nu}$ is not constrained to be symmetric.
Its antisymmetric part, the torsion tensor $S^\rho_{\phantom{\rho}\mu\nu}=\Gamma^{\rho}_{[\mu\nu]}$, is a variable in the principle of stationary action.
Square brackets denote antisymmetrization.
Regarding the metric and torsion tensors as independent variables gives the correct generalization of the conservation law for the total (orbital plus intrinsic) angular momentum to the presence of the gravitational field \cite{Niko,SK}.
Including torsion also removes the curvature singularity at the big bang and replaces it with a nonsingular bounce \cite{bb}, solves the flatness and horizon problems in cosmology \cite{infl}, and requires fermions to be spatially extended, which may provide an ultraviolet cutoff in quantum field theory \cite{non}.

We use the notations of Ref. \cite{Niko}.
The metricity condition $g_{\mu\nu;\rho}$, where the semicolon denotes the covariant derivative with respect to $\Gamma^{\rho}_{\mu\nu}$, gives $\Gamma^{\rho}_{\mu\nu}=\{^{\rho}_{\mu\nu}\}+C^\rho_{\phantom{\rho}\mu\nu}$, where $\{^{\rho}_{\mu\nu}\}=(1/2)g^{\rho\sigma}(g_{\mu\sigma,\nu}+g_{\nu\sigma,\mu}-g_{\mu\nu,\sigma})$ are the Christoffel symbols of the metric, $g^{\mu\nu}$ is the contravariant metric tensor, and $C^\rho_{\phantom{\rho}\mu\nu}=S^\rho_{\phantom{\rho}\mu\nu}+2S_{(\mu\nu)}^{\phantom{(\mu\nu)}\rho}$ is the contortion tensor.
Round brackets denote symmetrization.
The curvature tensor $R^\lambda_{\phantom{\lambda}\rho\mu\nu}=\Gamma^{\lambda}_{\rho\nu,\mu}-\Gamma^{\lambda}_{\rho\mu,\nu}+\Gamma^{\sigma}_{\rho\nu}\Gamma^{\lambda}_{\sigma\mu}-\Gamma^{\sigma}_{\rho\mu}\Gamma^{\lambda}_{\sigma\nu}$, where the comma denotes a partial derivative over coordinates, can be thus decomposed as $R^\lambda_{\phantom{\lambda}\rho\mu\nu}=P^\lambda_{\phantom{\lambda}\rho\mu\nu}+C^\lambda_{\phantom{\lambda}\rho\,\nu:\mu}-C^\lambda_{\phantom{\lambda}\rho\,\mu:\nu}+C^\sigma_{\phantom{\sigma}\rho\,\nu}C^\lambda_{\phantom{\lambda}\sigma\,\mu}-C^\sigma_{\phantom{\sigma}\rho\,\mu}C^\lambda_{\phantom{\lambda}\sigma\,\nu}$, where $P^\lambda_{\phantom{\lambda}\rho\mu\nu}$ is the curvature tensor constructed from the Christoffel symbols (the Riemann tensor) and the colon denotes the covariant derivative with respect to $\{^{\rho}_{\mu\nu}\}$ \cite{Niko,Schou}.
The Lagrangian density for the gravitational field in the Einstein--Cartan theory of gravity is the same as in the general theory of relativity, $\mathfrak{L}=-(1/2\kappa)R_{\mu\nu}g^{\mu\nu}\sqrt{-g}$, where $R_{\mu\nu}=R^\rho_{\phantom{\rho}\mu\rho\nu}$ is the Ricci tensor.
The Einstein--Cartan action is thus given by \cite{Niko}
\begin{eqnarray}
& & I=-\frac{1}{2\kappa c}\int R_{\mu\nu}g^{\mu\nu}\sqrt{-g}d\Omega=-\frac{1}{2\kappa c}\int g^{\mu\nu}(P_{\mu\nu}-2C^\rho_{\phantom{\rho}\mu\rho:\nu}-C^\rho_{\phantom{\rho}\mu\rho}C^\sigma_{\phantom{\sigma}\nu\sigma}+C^\rho_{\phantom{\rho}\mu\sigma}C^\sigma_{\phantom{\sigma}\nu\rho})\sqrt{-g}d\Omega \nonumber \\
& & =-\frac{1}{2\kappa c}\int g^{\mu\nu}(P_{\mu\nu}-C^\rho_{\phantom{\rho}\mu\rho}C^\sigma_{\phantom{\sigma}\nu\sigma}+C^\rho_{\phantom{\rho}\mu\sigma}C^\sigma_{\phantom{\sigma}\nu\rho})\sqrt{-g}d\Omega+\frac{1}{\kappa c}\int(g^{\mu\nu}C^\rho_{\phantom{\rho}\mu\rho}\sqrt{-g})_{,\nu}d\Omega.
\label{EC}
\end{eqnarray}
The last term in Eq. (\ref{EC}) can be written as a hypersurface integral,
\begin{equation}
I_\textrm{S}=\frac{2}{\kappa c}\int\mathfrak{g}^{\mu\nu}S_\mu df_\nu,
\end{equation}
where $\mathfrak{g}^{\mu\nu}=\sqrt{-g}g^{\mu\nu}$ is the contravariant metric tensor density and $S_\mu=S^\nu_{\phantom{\nu}\mu\nu}$ is the torsion vector.

Let us consider the variation of the action with respect to the torsion tensor.
Such a variation at the boundary of the integration domain is equal to
\begin{equation}
\delta I_\textrm{S}=\frac{2}{\kappa c}\int\mathfrak{g}^{\mu\nu}\delta S_\mu df_\nu=\frac{2}{\kappa c}\int\mathfrak{g}^{\mu 0}\delta S_\mu dV.
\end{equation}
Substituting this variation as $\delta I$ into Schwinger's principle (\ref{Schw2}) for $O=S_\mu$ gives
\begin{equation}
\delta S_\mu({\bf x},t)=-\frac{2i}{\hbar\kappa c}\biggl[S_\mu({\bf x},t),\int(\mathfrak{g}^{\nu 0}\delta S_\nu)({\bf x}',t)dV'\biggr],
\end{equation}
which is satisfied if
\begin{equation}
[S_\mu({\bf x},t),S_\nu({\bf x}',t)]=0
\label{com1}
\end{equation}
and
\begin{equation}
[S_\mu({\bf x},t),\mathfrak{g}^{\nu 0}({\bf x}',t)]=\frac{i}{2}\hbar\kappa c\delta^\nu_\mu{\bm\delta}({\bf x}-{\bf x}')=4\pi il^2_\textrm{P}\delta^\nu_\mu{\bm\delta}({\bf x}-{\bf x}'),
\label{com2}
\end{equation}
where $l_\textrm{P}$ is the Planck length.
The equal-time relations (\ref{com1}) and (\ref{com2}) are covariant under general coordinate transformations $x^\mu\to x'^\nu(x^\mu)$.

The quantum commutation relation (\ref{com2}) between the metric and torsion indicates that 1) the components $g^{\mu 0}$ of the metric tensor cannot be zero and 2) the torsion vector cannot be zero.
Accordingly, exact spherically and axially symmetric gravitational fields do not exist in the quantum theory.
In the classical Einstein--Cartan theory of gravity coupled to spinors, the only irreducible components of the torsion tensor that do not vanish are the components of the completely antisymmetric torsion tensor, so the torsion vector vanishes \cite{Niko,SK}.
The relation (\ref{com2}) therefore endows spacetime, even in vacuum, with intrinsic torsion.

Equation (\ref{com2}) shows that the mixed (time-space) components of the metric tensor are canonically conjugate to the torsion vector.
The analysis leading to this result is actually at the classical level.
Full quantization of the Einstein--Cartan theory must address the same problems (renormalizability and unitarity) that appear in quantization of general relativity.

\end{document}